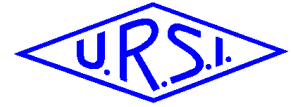

# The Relation between Type III Radio Storms and CIR Energetic Particles

Nat Gopalswamy*[1], Pertti Mäkelä[2], Seiji Yashiro[2], Sachiko Akiyama[2], and Hong Xie[2]
(1) NASA Goddard Space Flight Center, Greenbelt, MD 20771, USA, https://cdaw.gsfc.nasa.gov
(2) The Catholic University of America, Washington DC 20064, USA

## Abstract

We report on a study that compares energetic particle fluxes in corotating interaction regions (CIRs) associated with type III radio storm with those in non-storm CIRs. In a case study, we compare the CIR particle events on 2010 October 21 and 2005 November 2. The two events have similar solar and solar wind circumstances, except that the former is associated with a type III radio storm and has a higher CIR particle flux and fluence. We also perform a statistical study, which shows that the proton and electron fluences are higher in the storm-associated CIRs by factor of ~6 and 8, respectively than those in the storm-free CIRs.

## 1 Introduction

A high-speed stream (HSS) originating from a coronal hole (CH) forms a corotating interaction region (CIR) when HSS presses against the slow wind ahead. CIRs accelerate particles starting from tens of solar radii (Rs) from the Sun. When an active region (AR) is located near the edge of a CH, interchange reconnection (IRC) occurs between the CH open field (OF) and AR closed field. This OFAR configuration is required for the occurrence of a type III radio storm in the interplanetary medium [1-3]. The nonthermal electrons responsible for type III storms are thought to be accelerated in the IRC region. While ordinary type III bursts originate along open field lines near the main AR neutral line, the storm type III bursts occur on the AR periphery via IRC [4]. All known accelerators such as magnetic reconnection in current sheets (flares), shocks driven by coronal mass ejections, CMEs), and CIR shocks/compressions have been found to accelerate both electrons and ions. It is therefore logical to expect that IRC regions also accelerate ions in addition to accelerating electrons that produce type III storms. Type III storms originate close to the Sun (<2 Rs) [5]. The significance of this expectation is that particles accelerated in the IRC regions can serve as seed particles to the CIR accelerator. At tens of MHz, type I storms transition into type III storms [6]. Thus, type III storms represent a continuous drizzle of suprathermal electrons (2-100 keV) [7] that are likely to be accompanied by suprathermal ions.

In this paper we consider two CIR particle events, one accompanied by a type III storm, while the other is storm free to see if the particle events differ significantly. We also perform a statistical analysis of storm-free and storm-associated CIR particle events.

## 2 A Case Study

In this section we compare two CIR particle events, one associated with a type III storm and the other storm free, but the solar wind parameters are similar in the two events.

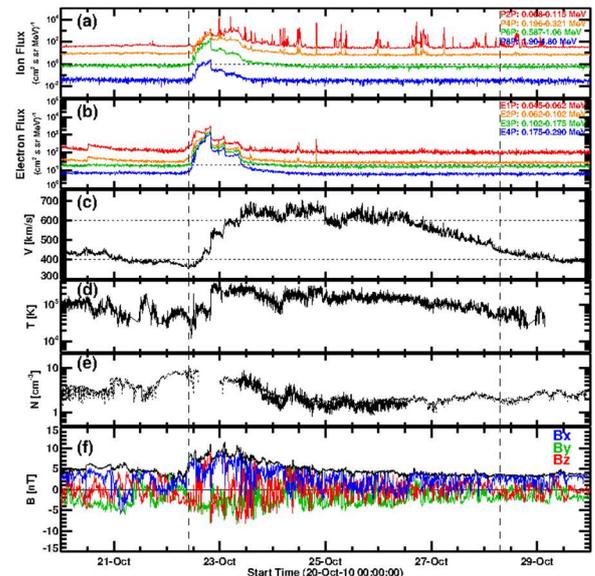

**Figure 1.** Solar wind conditions of the 2010 October 22-28 CIR event detected by ACE (between vertical dashed lines) associated with a type III storm. Ion and electron fluxes observed by the Low Energy Magnetic Spectrometer (LEMS) and Low Energy Foil Spectrometer (LEFS) on the Electron, Proton, and Alpha Monitor (EPAM), respectively. We have plotted the ion fluxes measured in 4 channels (out of the 8 LEMS channels) in the energy ranges, 0.068-0.115 MeV (P2P; red), 0.196-0.321 MeV (P4P; orange), 0.587-1.08 MeV (P6P; green), and 1.90-4.75 MeV (P8P; blue). Electron fluxes are also plotted in four energy channels: 0.045-0.062 MeV (E1P), 0.062 – 0.102 MeV (E2P), 0.102 – 0.175 MeV (E3P), and 0.175 – 0.290 MeV (E4P). The Solar wind speed (V), proton temperature (T), and proton density (N) are from ACE's Solar Wind Electron, Proton, and Alpha Monitor (SWEPAM) and the magnetic field with its components (B, Bx, By, Bz) plotted in GSE coordinate are measured by the Magnetometer (MAG).

## 2.1 A Storm-free CIR Particle Event

A CIR observed at Earth typically requires a coronal hole at the central meridian of the Sun about one day before the arrival of the stream interface and 3-4 days before the HSS reaches its maximum speed. If an active region is present at the edge of the coronal hole, an OFAR configuration results, required for type III storms. Figure 1 shows a CIR event starting on 2010 October 22. Both energetic electrons and ions are observed starting from the beginning of the CIR event, peaking at the interface, and decaying back to background values when the solar wind speed reaches its maximum along with a drop in proton density. We compute the particle fluences by summing up fluxes within the CIR interval that are in excess of the background. The proton fluence in the P6P channel (0.587 – 1.06 MeV) is $2.08 \times 10^6$ cm$^{-2}$ sr$^{-1}$ MeV$^{-1}$. The electron fluence in the E3P channel (0.102 - 0.175 MeV) is $2.20 \times 10^7$ cm$^{-2}$ sr$^{-1}$ MeV$^{-1}$.

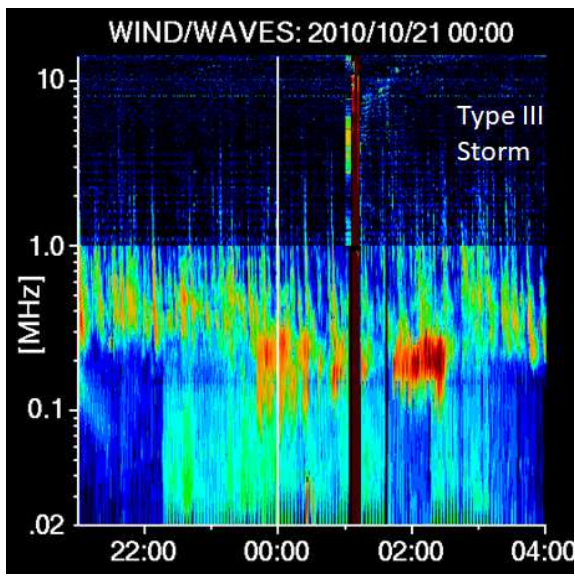

**Figure 2.** The 2010 October 21 type III storm observed by the Wind/WAVES instrument [8].

Figure 2 shows the type III storm observed by the Wind/WAVES instrument on 2010 October 11. The fine thread-like storm type III bursts can be seen extending down to about 0.2 MHz. The CH that results in the CIR is shown in Fig. 3a, flanked by two active regions (ARs) 11113 on the west and 11118 on the east. The OFAR configuration is possible on either side of the CH, but the one on the west is the likely source because the magnetic field configuration is conducive for IRC (positive side field lines of the AR adjacent to the CH negative field lines). Open field lines from the coronal hole intersect forward and reverse shocks that may form at tens of Rs from the Sun (Simnett and Roelof 1995) and these field lines are likely to carry electrons and ions accelerated at the edge of coronal holes where interchange reconnection takes place. The reverse shock propagating toward the Sun will have access to more and more field lines from the coronal hole.

## 2.2 A Storm-free CIR Particle Event

An example of a CIR with no accompanying type III storm is shown in Fig. 4. The interface is sharp in this event as indicated by the density and magnetic field peaks. The ion and electron fluxes increase above the background just before the density increase at the interface. The particle fluxes have a smooth time profile lasting for ~5 days, roughly following the solar wind speed. The particle fluxes decline with the solar wind speed. The electron and ion fluxes increase by only one order of magnitude in the CIR compared to the background. The duration of CIR particle enhancement is much longer than the 1.5 days in the 2010 October 22 event. However, the fluences are smaller than the corresponding values in the 2010 October 22 CIR event for protons ($1.74 \times 10^6$ cm$^{-2}$ sr$^{-1}$ MeV$^{-1}$) and electrons ($9.28 \times 10^6$ cm$^{-2}$ sr$^{-1}$ MeV$^{-1}$).

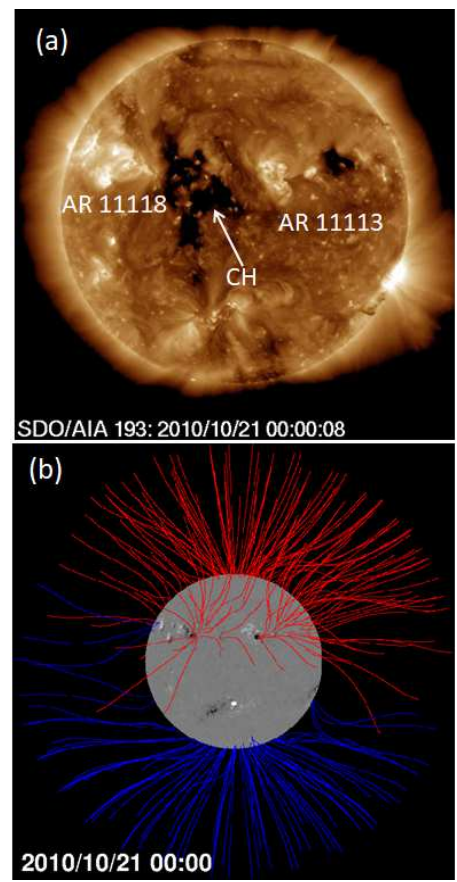

Figure 3. (a) The coronal hole and (b) PFSS open field lines (red: negative; blue: positive) on 2010 October 21. ARs 11113 and 11118 are indicated.

The Wind/WAVES dynamic spectrum in Fig. 5 shows that there is no type III storm in the 2005 November 2. The dynamic spectrum is very quiet except for a patch of auroral kilometric radio emission. The HSS responsible for the CIR in Fig. 4 originates from a coronal hole that rotates to the disk center on 2005 November 2 as can be seen in the SOHO/EIT image in Fig. 6a. The CH has negative polarity as indicated by the red open field lines from the disk center in Fig. 6b. There is a pair of ARs to the southeast of the CH. The negative patches of these ARs are located adjacent to the CH, so the configuration is not conducive for IRC. This is likely the reason there is no type III storm even though there is OFAR configuration.

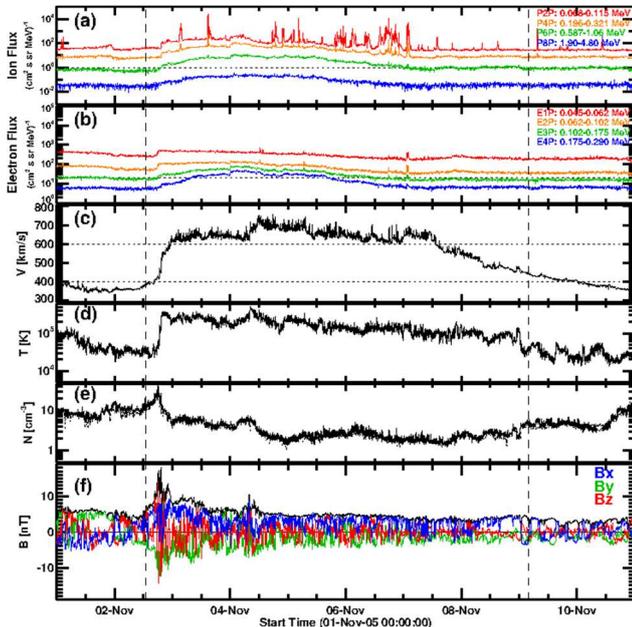

**Figure 4.** Same as Figure 1, but for the 2005 November 2 storm-free CIR particle event.

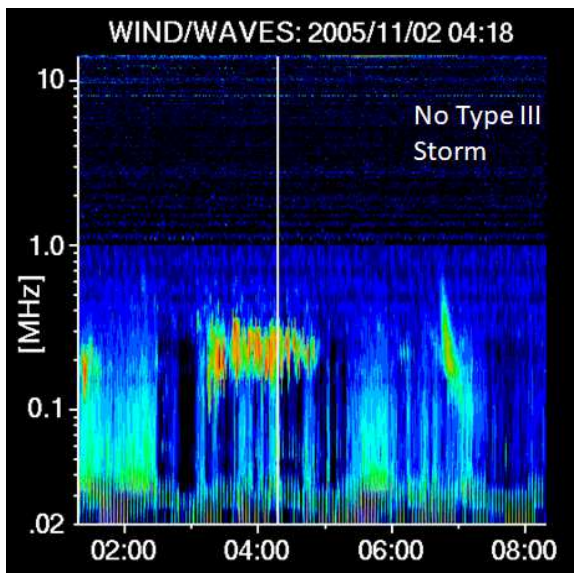

**Figure 5.** Wind/WAVES dynamic spectrum devoid of any type III storm on 2005 Nov 2. The intense emission marked by the white line near 4 UT is auroral kilometric radiation.

Comparing Figs. 1 and 4 we see that the underlying coronal holes and the high-speed stream (speed ~ 700 km/s) are similar in the two cases. The primary differences are higher CIR particle flux and the presence of type III storm in the case of the 2010 October 21 event. This result strongly suggests that the presence of the type III storm somehow influences the CIR particle flux.

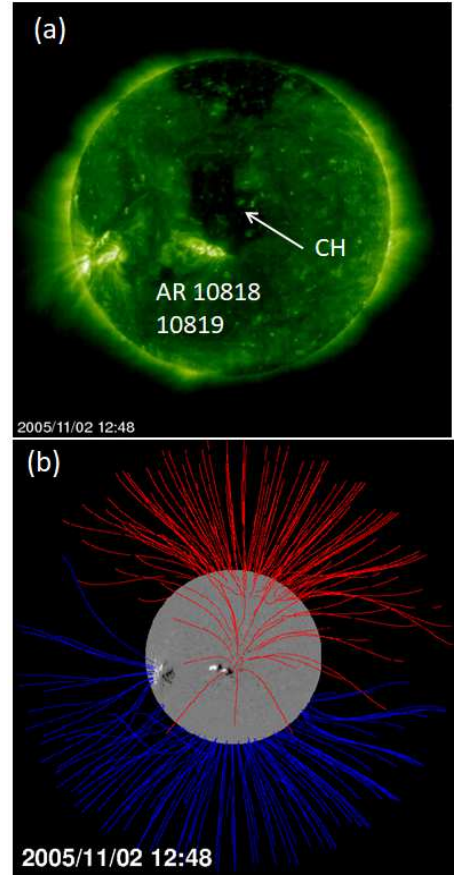

Figure 6. (a) The coronal hole (CH) and (b) PFSS open field lines (red: negative; blue: positive) on 2005 November 2. The bright patch at the southern edge of CH consists of ARs 10818 and 10819 at the left and right parts of the bright patch as is clear from the magnetogram (b).

## 3 A Statistical Study

In order to check the validity of the case study, we consider a large number of CIRs observed over solar cycles 23 and 24. We select 130 CIRs that are relatively isolated. Among the 130 CIR events, 18 are accompanied by type III storms; the remaining are storm-free. Following the procedure used in the case study we compute the fluences and compile the speeds of the high-speed streams in the 130 CIR events. Figure 7 shows the distribution of proton and electron fluences in the storm and storm-free events. It is clear that the proton and electron fluences have higher median values when they are accompanied by a type III storm. The dependence of the particle fluences on the solar wind speed is similar in storm-related and storm-free CIRs, but the storm-related events have a higher fluence. This suggests that the presence of type III storms makes the difference between the two sets of fluences. Thus, the case study presented in section 2 is supported by the statistical results (Fig. 7). This study shows that 18 out of 130 CIRs (or ~14%) are associated with type III storms. This study can be expanded to improve the statistics using a larger sample available in cycles 23 and 24 [9-10].

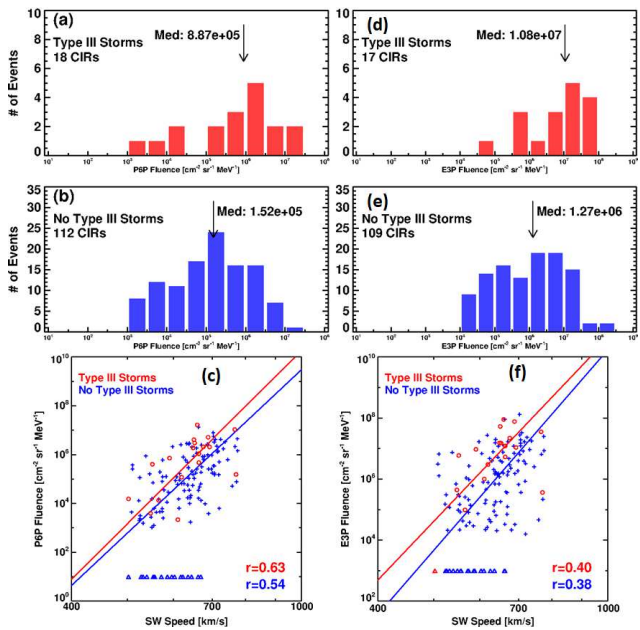

**Figure 7.** Proton and electron fluence distributions (a,d) in CIRs accompanied by a type III storm compared with the corresponding fluences (b,e) in storm-free CIRs. The median values of the distributions are noted on the plots. Correlations of proton (c) and electron (f) fluences with the solar wind speed. The triangles are events in which the fluences are below the threshold ($<10^3$ cm$^{-2}$ sr$^{-1}$ MeV$^{-1}$ for protons and ($<10^4$ cm$^{-2}$ sr$^{-1}$ MeV$^{-1}$ for electrons). The listed correlation coefficients are significant because the Pearson's critical coefficient are 0.4 and 0.165 for the storm and storm-free subsets of CIR events.

## 4 Discussion and Summary

The primary finding of this paper is that a CIR particle event is more intense when it is accompanied by a type III storm. Type III storms are produced by nonthermal electrons accelerated near the Sun in current sheets formed between the closed field lines of active regions and the open field lines in the adjacent coronal holes. CIR particles are accelerated at tens of Rs from the Sun in the interaction region [11-13]. What is the connection between the two phenomena? We suggest that electrons and protons accelerated in the interchange reconnection region of the OFAR magnetic configuration travel to the CIR along the open field lines in the coronal hole and serve as seed particles to the CIR accelerator. These seed particles are in addition to the suprathermal tail of the solar wind thought to be the seed population for all CIRs. Now that CIR particles [12] and type III storms [14] are observed closer to the Sun using Parker Solar Probe, we have the opportunity to further explore the connection between type III storms and CIR particle populations. Type III storms as a source of seed particles have not been considered before, although many types of seed particles are considered [15].

## 5 Acknowledgements

We benefited from the open data policy of *SOHO*, *STEREO*, *SDO*, *GOES*, *ACE*, and *Wind* teams. Work supported by NASA's LWS program.